\documentclass[
    ,final            
  ]
  {aipproc}

\layoutstyle{8x11double}

\begin{document}

\title{Central Compact Objects in Supernova Remnants}


\author{Andrea De Luca}{
  address={IASF/INAF Milano, Via Bassini 15, 20133 Milano, Italy}
  ,altaddress={IUSS Pavia, V.le Lungo Ticino Sforza 56, 27100 Pavia, Italy} 
}

\begin{abstract}
Central Compact Objects (CCOs) are a handful of soft X-ray sources
located close to the centers of Supernova Remnants and supposed to 
be young, radio-quiet Isolated Neutron Stars (INSs). A clear understanding of their
physics would be crucial in order to complete our view of the birth properties
of INSs. We will review the phenomenologies of CCOs, underlining the most 
important, recent results, and we will discuss the possible relationships of
such sources with other classes of INSs.
\end{abstract}


\maketitle

\section{CCOs \& the many species of INSs}
Recent X-ray observations radically changed the classic
idea that all Isolated Neutron Stars (INSs) are born as fast spinning radio
pulsars. A rich phenomenology emerged, which led to the classification
of INSs into different species. Radio-loud species include rotation-powered
radio PSRs 
and Rotating Radio Transients (RRaTs, \citep{mclaughlin06}). 
The other species are generally 
radio-quiet and include Anomalous X-ray Pulsars (AXPs, Woods,
these proceedings; see also \citep{woods06}), Soft Gamma 
Repeaters (SGRs, Woods, these proceedings; see also 
\citep{woods06}), Central Compact Objects (CCOs, discussed
here; see also \citep{pavlov02,pavlov04}) at the center of Supernova
Remnants (SNRs) and
X-ray Dim Isolated Neutron Stars (XDINSs, Kaplan, these proceedings;
see also \citep{haberl06}). 

The differences among the INSs' species
are certainly related to different 
properties of their magnetic fields. For instance,
AXPs and SGRs are supposed to be close relatives, different
from standard radio PSRs owing to their huge 
magnetic field (hence named ``magnetars'').
Unifying the rich phenomenological diversity in a coherent physical scenario
is one of the most urgent tasks in INS astronomy. 
A clear picture, including evolutionary paths, 
possibly connecting different species, is still lacking.

To this aim, understanding the birth properties of INSs would be crucial.
Indeed, 
the least understood members of the INSs family are the youngest ones,
i.e. the CCOs.

CCOs (see Table 1 for a list of the seven members of the class)
are a handful of sources characterized by (i) position close to the 
center of a young SNR; (ii) lack of radio/IR/optical counterparts,
as well as of surrounding diffuse, non-thermal nebulae; 
(iii) constant, unpulsed X-ray emission ($L_X\sim10^{33}$ erg s$^{-1}$)
with thermal-like spectrum characterized by high temperatures
(0.2-0.4 keV) over a very small emitting area (few \% of the expected 
NS surface).
Thus, while point (ii) implies that CCOs are not standard
young radio PSRs, point (iii) separates them from ``standard'' AXPs and
SGRs. The classification of a source as a CCO has been in some way a process
by elimination, in the lack of a clear physical understanding of such sources.

We are not even sure that all CCOs be INSs. We only know for sure that they
are young and that
their formation in supernova explosions
must be a rather common event. Indeed, inspecting all known SNRs within 5 kpc
of the solar system, we find 14 radio PSRs (3 are beamed away from us, revealed
by bright non-thermal nebulae), 6 CCOs and 1 AXP. New candidate
CCOs are also being discovered at the centers of more distant SNRs.

Recently, results on specific sources shed light on their nature.
We will review such new results
and we will exam possible classification schemes.

\begin{table}
\begin{tabular}{c c c | c c c c c}
\hline \hline
SNR & Age & Distance & Observed flux & Luminosity & Variability &
Period & Pulsed fraction \\
 & (ky) & (kpc) & $10^{-12}$ erg cm$^{-2}$ s$^{-1}$ & $10^{33}$ erg s$^{-1}$ &
& & \\
\hline
RCW103      & $2$      & $3.3$      & 0.8-60 & 1.1-80 & factor 100      & 6.67 hour & 12-50\%  \\
G296.5+10.0 & $7$     & $2.2$      & 2.     & 1.2 & $<5\%$   & 424 ms    & $\sim10\%$ \\
Kes 79      & $7$      & $7.1$      & 0.2    & 3 & $<15\%$  & 105 ms    &
$\sim80\%$\\
Cas A       & 0.3      & $3.4$      & 2.     & 2 & Flares?  & ...       & $<13\%$ \\
Puppis A    & $3.7$    & $2.2$      & 4.8    & 5 & $<5\%$   & ...       & $5\%
?\,\,(<7\%)$ \\
G347.3-0.5  & $2$      & $1.3$      & 3.     & 0.6 & $<5\%$   & ...       & $<7\%$ \\
VelaJr.     & $1$      & $1$        & 1.3    & 0.25 & $<5\%$   & ...       & $<7\%$ \\
 \hline
\end{tabular}
\caption{List of the seven ``confirmed'' CCOs and of their basic X-ray
  properties. Flux is in the 0.5-8 keV energy range; the bolometric luminosity
  is computed for a purely thermal model (either single or double blackbody). 
See text for references.}
\end{table}

\section{Searching for a CCO template}
\subsection{1) A very peculiar magnetar}
1E 161348-5055 (1E 1613) was discovered with the Einstein satellite
\citep{tuohy80} 
very close to the geometrical center of the young \citep[2,000 yr, ][]{carter97} 
supernova remnant RCW103, located at a distance of $\sim3.3$ kpc 
\citep{reynoso04}. 

Historically, it was the first radio-quiet neutron star candidate
found inside a SNR.
While 1E 1613 was considered to be most probably
a young, off-beamed pulsar, the lack of a surrounding, diffuse non-thermal
nebula \citep{tuohy80} made it very different from the Crab, the prototypical young pulsar.
Such an interpretation was supported by
its soft X-ray spectrum, pointing to the first
detection of thermal radiation from the surface 
of a NS, as well as by the lack of a radio or 
optical counterpart \citep{tuohy83,gotthelf97}.

However, X-ray observations of 1E 1613 over the following years
unveiled a puzzling temporal behaviour.
\citet{gotthelf99}, 
using
ASCA, ROSAT and Einstein data, found evidence for a factor 10
variability in flux on the few year time scale. More recently,
such a variability was confirmed thanks to Chandra observations.
\citet{garmire00} discovered a large brightening (about two orders
of magnitude) between September 1999 and March 2000, while two years
later \citet{sanwal02} 
(with Chandra) as well as \citet{becker02} (with XMM-Newton) observed 1E 1613
at an intermediate flux level. Moreover, the first Chandra observation
of 1E 1613 in its low state
hinted a possible periodicity at $\sim6$ hours \citep{garmire00b}.
The subsequent observations of the source in active state
could not conclusively solve the issue, the periodicity was not detected
in the very high state of early 2000, but was possibly seen again in 2002 by
\citet{sanwal02}, while \citet{becker02} did not find any 
periodicity, but observed a complex light curve including a 
possible ``partial eclipse''. 

A long (90 ks) observation 
with XMM-Newton, performed in 2005, caught 1E 1613 in a low state 
and yielded unambiguous evidence for a strong, nearly sinusoidal
modulation at P=6.67$\pm$0.03
hours, with a 50\% pulsed fraction \citep{deluca06}. 
The source spectrum,
well described by an absorbed double blackbody model, varies
significantly
as a function of the 6.67 hour cycle and appears harder 
at the peak.  The same 6.67 hour periodicity was 
then recognized also in the older XMM-Newton dataset, with a much lower 
pulsed fraction ($\sim10\%$) and a remarkably different light curve,
including two narrow minima (``dips'') per period. Such an
``active state'' was also characterized by a factor 6 higher flux
and a harder spectrum.
No faster pulsations are seen in the two XMM-Newton observations 
down to P=12 ms, 
with an upper limit of 10\% to the pulsed fraction (at 99\% c.l.).

Monitoring with Swift/XRT shows that the source (as of August, 2007)
is still fading, although at a somewhat slower rate. A long 
observation with Chandra/HRC,
performed in 2007, July by our group, shows again the nearly sinusoidal
modulation at 6.67 hours, with a pulsed fraction as high as $\sim55\%$
in 0.1-10 keV.

On the optical/IR side, VLT/ISAAC and HST/NICMOS images 
collected in 2001 and in 2002, respectively, unveiled
a very crowded field, with a few objects possibly consistent
with the X-ray position of 1E 1613 \citep{sanwal02,mignani04,pavlov04}.
Deep observations with the VLT/NACO instrument were performed in 2006
(during the low state of 1E 1613) with the aim to search for 6.67 hours 
modulation of the four possible counterparts (Ks$\sim18-20$)
lying within the $3\sigma$ error region (Mignani et al., these proceedings; 
\citep{deluca07a}). None was found.
Comparison with the HST/NICMOS
images does not show a clear variability correlated to the factor 3.5 fading
of 1E 1613 during the same time span. Moreover, the possible counterparts do not
stand out for peculiar colors with respect to the bulk of the very red
(H-K$\sim2$, requiring A$_V\sim20-25$) stellar background population.
Thus, there are no compelling reasons to associate any of them to 1E 1613, which
remains undetected in the IR down to Ks$\sim22.1$. A search for 
a counterpart with Spitzer was also performed, with negative results \citep{wang07}.

\begin{table}
\begin{tabular}{c|c|c}
\hline \hline
CCO & optical/IR counterpart & optical/IR upper limit\\
 & (mag) & (mag) \\
\hline
1E 1613 in RCW103      & H$\sim21.4\,(?)$, K$\sim19.2\,(?)$ & I$>25$, Ks$>22.1$ \\
1E 1207 in G296.5+10.0 & J$\sim21.7$, H$\sim21.2$, K$\sim20.7$ & R$>27.1$, J$>23.5$, H$>22.4$, K$>22.0$ \\
CXOU J1852 in Kes 79      & ... & R$>24.9$ \\
CXOU J2323 in Cas A       & ... & R$>27.8$, J$>26.2$, H$>24.6$, K$>21.2$ \\
RX J0822 in Puppis A    & ... & B$>26.5$, R$>26.0$, J$>21.7$, H$>20.6$, Ks$>20.1$ \\
1WGA J1713 in G347.3-0.5  & H$\sim19.4\,(?)$, Ks$\sim18.3\,(?)$ & I$>24.6$, H$>22$, Ks$>20.5$ \\
CXOU J0853 in VelaJr.     & H$\sim21.6\,(?)$, Ks$\sim21.4\,(?)$ & R$>25.6$, J$>22.6$, H$>22.5$, Ks$>21.8$ \\
 \hline
\end{tabular}
\caption{Optical/infrared results for the seven confirmed CCOs. 
In the case of 1E 1207 we give
the magnitudes of an M dwarf located close to the Chandra position; updated
astrometry questioned its possible association to 1E 1207 \citep{fesen06,wang07}. 
A few IR
sources have been found inside the Chandra error circle for the CCOs in
the RCW103 and G347.3-0.5 SNRs. In such cases, the magnitudes refer to the 
source closest to the X-ray position, even if there are not
compelling reasons to associate the IR sources to the 
X-ray ones. 
See text for references.}
\end{table}

\subsubsection{Is 1E1613 a ``braked magnetar''?}

The unique combination of 6.67 hour periodicity, dramatic long-term
variability, young age and underluminous IR counterpart makes 1E 1613 a unique
source among all compact objects. 

Association of 1E 1613 to RCW103 seems very robust, based on several
arguments. The point source lies within 15 arcsec of the apparent
center of the 10 arcmin wide SNR. Moreover, the two system have
consistent distances, as apparent by the same interstellar X-ray
absorption \citep{deluca06}, as well as by neutral H studies in radio,
which also support the association on a morphological basis \citep{reynoso04}.

As discussed by \citet{deluca06},  1E 1613
could be a binary system featuring a compact object, born in the 
supernova event which generated RCW103, and a very small companion star.
In such a frame, the 6.67 hour modulation could be ``naturally'' 
interpreted as the orbital period of the system. However, 1E 1613 is
dramatically different from any known Low-Mass X-ray Binary  (LMXB) system,
because of its low luminosity ($10^{33}-10^{35}$ erg s$^{-1}$),
purely thermal spectrum, large spectral evolution
along the 6.67 hour cycle, long-term variability in pulse shape
and fraction, very long time scale for the recovery from the outburst.
\citet{deluca06} proposed that a peculiar ``double accretion'' (wind + disc)
scenario could be at work in a very young LMXB, driven by a
significant orbital eccentricity, expected on theoretical basis \citep{kalogera96}. 
The recent IR results do not support such a picture \citep{deluca07a}. 
None of the potential
counterparts is consistent with a small star at the distance of 1E 1613. 
The upper limits leave room only for a very low-mass
star (M6-M8), which seems unable to power via its wind the observed
pulsed luminosity (an accretion rate of $\sim10^{-13}\,M_{\odot}\,yr^{-1}$
would be required). 
Moreover, it seems unlikely that a LMXB with such an extreme 
mass ratio could survive the supernova explosion.
Such difficulties\footnote{A different binary picture for 1E 1613, suggesting 
the system to be an analog of Cataclysmic Variables of the Polar
or Intermediate Polar classes, originally proposed by \citet{popov07}, has been studied
by \citet{pizzolato07}. Such a scenario, which could possibly avoid same of our 
drawbacks,  features a magnetar in a
binary system with a low-mass star. Magnetic and material interaction
could have slowed down the NS rotation to P=6.67 hours, synchronous
(as in Polars) or quasi-synchronous (as in Intermediate Polars) to
the orbital period.} lead us to consider an alternative picture of 1E 1613
as a very peculiar isolated compact object \citep{deluca06}.
Within such a frame, the picture best fitting to the unique phenomenology of
1E 1613
is the one of a ``braked magnetar'', spinning at 6.67 hours.
Indeed, most aspects of 1E 1613's phenomenology easily fit in a magnetar scenario:
spectrum, luminosity, long term variabilities are very similar to the ones
shown by Anomalous X-ray Pulsars \citep{woods06}.
However, all known AXPs spin in the 2-12 s range, i.e. thousands of times faster
than 1E 1613. A very efficient braking mechanism is required to slow down 1E 1613 in
2000 yr from its presumably much faster spin rate at birth. \citet{deluca06} show
that propeller effect on the material of a fallback disc could provide such 
a mechanism, provided that the NS was born with a very high magnetic field
($\sim5\times10^{15}$ G) and with a rather slow period ($\sim300$ ms) to avoid an
early ``ejector'' phase which could have pushed away any surrounding
material. Recently, \citet{li07}, using a different model for
the interaction between the rotating INS's magnetosphere and the
surrounding fallback disc, showed that initial conditions may be 
relaxed and birth period down to a few ms could be allowed.
Thus, 1E 1613 would be the first known example of a new class of very
slowly rotating magnetars, whose spin down history is completely
dominated by the role of fallback material.

\subsection{2) Weakly magnetized INSs}
\subsubsection{1E 1207.4-5209 in G296.5+10.0}
1E 1207.4-5209 (1E 1207) was detected with the Einstein satellite close
to the center of the $\sim7$ kyr old SNR G296.5+10.0
\citep{roger88}, located at
a distance of $\sim2$ kpc
\citep{giacani00}, quite high to the Galactic Plane (b$\sim10^{\circ}$).
It was the 
second thermally-emitting radio-quiet INS candidate found inside a SNR. 
Pulsations from 1E 1207 were discovered with the Chandra satellite 
\citep{zavlin00}, proving the source to be an INS. 


Early timing investigations hinted a non-monotonous period 
evolution of 1E 1207, suggesting that the source could be a
peculiar binary system
\citep{zavlin04,woods07}. 

However, very recently, \citet{gotthelf07}, using at once
all available X-ray data, provided conclusive evidence that
1E 1207 is a very stable rotator, with essentially no measurable
period evolution
(see also Gotthelf \& Halpern, these proceedings). 
The upper limit to the period derivative 
($\dot{P}<2.5\times10^{-16}$ s s$^{-1}$ at 2$\sigma$) yields
an INS carachteristic age $\tau_c>27$ Myr, exceeding by 3 orders of magnitude
the age of the SNR, and a very small dipole magnetic field,
B$<3.3\times10^{11}$ G. Such results point to a
weakly magnetized INS, born with a spin period very close to the current one.

\subsubsection{CXOU J185238.6+004020 in Kes 79}

A very similar picture emerged for another member of the CCO class.
The source CXOU J185238.6+004020 (CXOU J1852) 
was discovered with a Chandra observation
by \citet{seward03} 
at the center of Kes 79 SNR, a 5.5-7.5 kyr old SNR \citep{sun04}, 
located at $\sim7$ kpc.
A follow-up observation with XMM-Newton allowed \citet{gotthelf05}
to discover a 105 ms pulsation from the source. Further observations 
with XMM-Newton and Chandra did not show a significant change in the 
period of CXOU J1852. \citet{halpern07} set a 2$\sigma$ upper limit 
to the period derivative $\dot{P}<2.0\times10^{-16}$ s s$^{-1}$,
yielding a characteristic age $\tau_c>8$ My and a dipole magnetic
field B$<1.5\times10^{11}$ G.

\subsubsection{Half-brothers or twins$?$}
Judging on the basis of their very similar spin
parameters, 
1E 1207 and CXOU J1852 should be close relatives. However, their 
spectra, as well as their phase-resolved behaviour, are very
different.

1E 1207 stands out among CCOs because of its unique spectrum.
Two large absorption features superimposed to the thermal spectrum,
centered at 0.7 keV and at 1.4 keV,
were discovered thanks to Chandra and XMM-Newton observations
 \citep{sanwal02,mereghetti02}.
Such features vary as a function of the rotational
 phase \citep{mereghetti02}. 
This was the first detection of spectral features in the X-ray
 spectrum of an INS,
making 1E 1207 an outstanding source among all compact objects.
A very deep (250 ks) observation performed with XMM-Newton in 2002
unveiled the presence of
a third absorption feature at 2.1 keV and possibly of a fourth one at 
2.8 keV \citep{bignami03,deluca04}. The actual significance of the third 
and fourth lines has been questioned by \citet{mori05} who 
evaluated the dependence of such two features' equivalent width on the
underlying continuum model.
The very deep XMM dataset of 2002 showed that 
the 424 ms $\sim7\%$ pulsation is almost entirely due to phase variation of the 
absorption features (with a $\sim12\%$ variation), while the continuum 
has a much less pronounced modulation ($\sim3\%$). Such
a behaviour is unique among all INSs.

The nature of the spectral features of 1E 1207 has been debated since
their discovery, possible interpretations being atomic transition
lines in the NS atmosphere or cyclotron features in the plasma
surrounding the star \citep{sanwal02,mereghetti02}. The cyclotron 
interpretation was strongly supported by the detection of 
the third and of the possible fourth line \citep{bignami03,deluca04},
since the four features have central energies in the harmonic ratio 1:2:3:4
and show a significant dependence on the NS rotational 
phase. Assuming the 0.7 keV feature to be the fundamental cyclotron line
yields a measure of the magnetic field of $8\times10^{10}$ G, or
$1.6\times10^{14}$ G, in case electron or protons be responsible 
for the absorption, respectively.
The scenario of 1E 1207 as a weakly magnetized neutron star
is fully consistent with the electron cyclotron
interpretation of the features. A difficulty with the cyclotron
scenario is posited by the similar equivalent widths 
observed for the first and second harmonic, at odds with theoretical expectations,
since the oscillator strength of the second harmonic should be
a factor $\sim2,000$ lower than the one of the first harmonic.
A possible solution to such a problem was proposed by \citet{liu06},
who suggested the magnetized plasma responsible for the absorption 
to be optically thick at the frequency of the first harmonic (so that 
a saturation absorption is achieved, independent on the particle density),
but optically thin for the second and higher harmonics. Such a model 
requires a rather high particle column density in the surroundings of
the NS, which could possibly be sustained by accretion of fallback material.
Alternative interpretation for the lines are also proposed (Ho et al., 
these proceedings).

CXOU J1852, on the other side, has a thermal spectrum with no features
within the statistics available, which is far less abundant than that for 1E 1207.
However,
CXOU J1852  has a striking peculiarity, i.e. it has
a very large pulsed fraction, as high as $\sim80\%$. Such a value
makes CXOU J1852 an outstanding source among all thermally emitting INS.
Such a phenomenology would point to the picture of a small hot
region on the NS surface, coming into view or being hidden as
a function of the star rotation. However, this is quite at odds
with the  picture of CXOU J1852 as a weakly magnetized NS. 
First, the small magnetic field inferred from
timing does not seem able to generate such a large surface 
temperature anisotropy (either due to anisotropic thermal conduction
from the stellar interior, or due to surface bombardment by magnetospheric 
particles). Moreover, gravitational bending of the trajectories of photons
escaping from the surface should significantly suppress the modulation. 
Indeed, \citet{psaltis00} showed that a pulsation larger than $\sim35\%$
cannot be expected even for an extremely small hot spot with a very
large temperature contrast with respect to the surface. Beaming due to
radiative transfer effects in a strongly magnetized plasma could explain
the modulation, but would require presence of large multipole components
in the magnetic field, in order to be consistent with the observed small 
spin-down. Alternatively, such problems could be solved in a picture
invoking accretion of fallback material. Emission coming from a region
related to an accretion column, possibly located at some heigth from the star surface,
could account for both the small emitting area and the high modulation.

\subsection{3) A dormant magnetar ($?$)}
With an age of $\sim330$ yr, as estimated with a HST study
of the expansion of high-velocity debris \citep{fesen06}, 
Cas A is the remnant of the last supernova explosion occurred in our Galaxy.
Detection of O and Si-group abundances in the ejecta supports the picture of
Cas A as the remnant of a massive star \citep{chevalier78}. 
The central X-ray source, CXOU J232327.9+584843 (CXOU J2323)
was discovered in the Chandra First-light image
\citep{tananbaum99} and identified a posteriori in ROSAT and Einstein images.
It lies $\sim7$ arcsec off the apparent SNR expansion center \citep{fesen06},
implying a (projected) velocity of order 350 km s$^{-1}$.
Extensive multiwavelength observations of both the CCO and the SNR have been
performed \citep{pavlov00,chakrabarty01,mereghetti02b,fesen06} and
different hypotheses (either an INS, or an iolated black hole) have
been considered to explain the CCO. 

A very interesting result was obtained by \citet{krause05} who
discovered in multi-epoch Spitzer images (at 24 $\mu \,m$), spanning a 1 year time interval,
fast moving features ($10-20$ arcsec yr$^{-1}$) located in the outskirts of the
SNR. At the SNR distance, such proper motion corresponds to a velocity close to c.
The most likely interpretation of such features is that they are infrared
echoes from interstellar dust, heated by a travelling pulse of light. This points
to a large flare from CXOU J2323, occurred around A.D. 1953, with an almost
orthogonal beaming with respect to the line of sight,
with a luminosity
of $\sim2\times10^{46}$ erg s$^{-1}$), which is comparable to the energetics 
of giant flares from SGRs.
If such an interpretation is correct, CXOU J2323 could be a dormant magnetar.
The spectrum and luminosity are consistent with that of of transient AXPs
in quiescence, as well as with SGRs observed in low-luminosity state 
\citep{mereghetti06}.
Current upper limits to long-term variability and pulsations in the soft X-ray
band \citep{mereghetti02b,fesen06b}, 
as well as upper limits to an IR counterpart \citep{fesen06b,wang07} are consistent
with such an hypothesis.

\section{Other CCOs: more of the same$?$}
\subsection{The central source in Puppis A}
Puppis A is the remnant of the explosion of a very massive star
\citep{canizares81}, occurred $\sim3,700$ years ago \citep{winkler88},
at a distance of $\sim2.2$ kpc \citep{reynoso03}. 
The central X-ray source RX J0852.0-4622 (RX J0852), hinted in Einstein images
\citep{petre82} and later identified with ROSAT \citep{petre96}, is
located $\sim6.1$ arcmin off the geometrical center 
of the SNR. Association of RX J0852 to the SNR is supported by consistent
distance estimates and HI morphological studies in radio
\citep{reynoso03}. Deep radio observations set very stringent upper 
limits to a radio nebula associated to RX J0852 \citep{gaensler00}.
The large offset between  RX J0852 and the SNR center 
requires a high space velocity for 
the compact object, inherited from a natal kick during the supernova
explosion. Indeed, evidence for a large proper motion in good
agreement with the expected one (both in direction and in magnitude)
has been reported, based on the analysis of multi-epoch Chandra images
\citep{hui06,winkler07}. 

Analysis of two XMM-Newton datasets did not confirm
a pulsation at 75 ms hinted in ROSAT data \citep{pavlov99} -
excluded also by Chandra data \citep{pavlov02} -
but yielded some evidence for a candidate periodicity around 220 
ms \citep{hui06b}, with a pulsed fraction of $\sim5\%$.
However, the significance of such a pulsation in each dataset is
rather low, and the corresponding periods at the two epochs
are rather different, which would imply a very large
period derivative ($\sim2\times10^{-10}$ s s$^{-1}$),
among the largest ever observed for an INS,
only comparable to the upper side of the values measured for an
extreme object such as SGR 1806-20 \citep{woods06}. The resulting characteristic 
age of $\sim17$ yr
would also require a non-steady spin down for the source.
New observations are needed to confirm (or to rule out)
such a peculiar periodicity.
We estimate that the currently available photon statistics
should allow to detect at 99\% confidence level any
modulation with a pulsed fraction higher than 7\% and
period in the range 12 ms - 20 s. Such a value (computed
assuming a sinusoidal modulation and accounting for the 
number of trials)
may be assumed as an upper limit to any undetected pulsation.

Spectrum and luminosity of the CCO, as seen by XMM-Newton,
are fully comparable to those of the other members of the family
\citep{hui06b}. No variability is apparent on the few month time scale,
with an upper limit of order 5\%. Upper limits to an optical/IR 
counterpart leave room for a faint dwarf star as well as for 
a fallback disc \citep{wang07}.

\subsection{The central source in G347.3-0.5}

The supernova remnant G347.3-0.5 is the prototype of the
peculiar class of ``non-thermal'' SNRs. 
Very faint in radio, and dominated, in the soft X-ray
band, by non-thermal emission \citep{koyama97,slane99}, 
the SNR is very bright 
at TeV energies, where it has been beautifully resolved in HESS images
\citep{aharonian04}. The distance and age of the remnant are debated.
A distance of order 6 kpc has been assumed in the past, based on a possible 
association of the SNR with surrounding molecular clouds and HII region 
\citep{slane99}. Such a distance would imply an age of a few $10^4$ yr,
assuming Sedov evolution. However, more recently, studies with XMM-Newton,
coupled to new CO mm-wave high-resolution observations,
unveiled a possible interaction of the SNR shock with molecular gas,
pointing to a distance of $1.3\pm0.4$ kpc \citep{cassim04,cassim04b,fukui03}. The 
revised distance implies a much younger age for the SNR (few thousands yr),
in agreement with the idea that G347.3-0.5 could be the remnant
of the supernova recorded in A.D. 393 \cite{wang97}.

The central X-ray source 1WGA J1713.4-3949 (1WGA J1713)
was observed by ROSAT \citep{pfefferman96,slane99}
and ASCA \citep{slane99}. 
XMM-Newton and Chandra observations confirmed its similarity
to other CCOs, on the basis of its thermal-like spectrum and
of the lack of any counterpart \citep{lazendic03,cassim04b}. 
At the revised SNR distance, the luminosity of 1WGA J1713 is
fully consistent with that of other members of the CCO class.

Our analysis of multi-epoch XMM-Newton observations does not show
any long-term flux variability larger than 
$\sim5\%$ on years time scale, nor pulsations with pulsed fraction larger than 
$\sim7\%$ in the 12 ms - 6 s range (at 99\% confidence level, taking 
into account the number of trials). 

In the optical/IR range, observations with VLT/NACO have been performed
in the H and K band (Mignani et al., these proceedings; \cite{mignani07}). 
A few faint sources ($Ks\sim18-19$) in a very crowded field are
possibly consistent with the Chandra position; however, no firm 
conclusions may be drawn about their association with 1WGA J1713.

\subsection{The central source in Vela Jr.}

The supernova remnant was discovered in ROSAT data, superimposed to
the large Vela SNR and emerging at energies above $\sim1$ keV \citep{aschenbach98}.
It is dubbed ``Vela Jr.'' because of its supposedly younger age
than the surrounding Vela remnant.
Indeed, the age and distance of the Vela Jr. SNR are a matter of controversy. 
Possible detection with Comptel 
of $\gamma$-ray line emission at 1.157 MeV - originating
from the decay of $^{44}$Ti produced in the SN explosion - suggested
a very young age ($<700$ yr) and small distance ($\sim200$ pc)
for the remnant \citep{iyudin98}. However, reanalysis of Comptel data questioned the
significance of the 1.157 MeV feature \citep{schonfelder00}. 
A possible emission feature
at 4.4 keV, detected at rather low significance ($\sim4\sigma$) in XMM-Newton data
\citep{iyudin05} and hinted in ASCA data \citep{tsunemi00} (but see also
\citep{slane01}), possibly due to $^{44}$Sc and $^{44}$Ti
fluorescence, 
supported the picture of a very young and nearby system. 

On the other side, the 
observed X-ray interstellar absorption is a factor $\sim6$ larger than 
the one observed towards the Vela SNR,  
arguing for a significantly larger distance 
for Vela Jr. \citep{slane01}. Considering all uncertainties,
a distance in the range 0.5-1.5 kpc and
an age in the range 1000-3000 yr seem reasonable estimates.
Vela Jr. is another member of the class of non-thermal SNRs.
It has a purely non-thermal soft X-ray emission
\citep{tsunemi00,slane01}, and it
has been detected at TeV energies \citep{aharonian05}. 
Thus, it appears very similar to G347.3-0.5, 
considering the fact that both sport
a CCO close to their center.

The central source, hinted by ROSAT images \citep{aschenbach98},
was observed in BeppoSAX data \citep{mereghetti01} and was finally
localized with high accuracy with Chandra \citep{pavlov01}. 
The lack of an optical counterpart points to an 
INS nature. The CCO, CXOU J085201.4-461753 (CXOU J0852),
is located $\sim4$ arcmin North wrt. the geometrical center of the 
SNR. The region is rather complex and radio observations
yield evidence for a diffuse source (possibly a planetary
nebula) very close (in projection) to the position of CXOU J0852
 \citep{reynoso06}. 

CXOU J0852 has been repeatedly observed with XMM-Newton \citep{becker06}
and Chandra \citep{pavlov01,kargaltsev02}. It has a thermal featureless spectrum
and a luminosity of $\sim2.5\times10^{32}$ erg s$^{-1}$ (at 1 kpc), the smaller among
the CCO group. 
Our analysis of the entire XMM
dataset allows to set an upper limit of 5\% to any long-term variability,
as well as an upper limit of 7\% to the pulsed fraction of 
any undetected pulsation in the range 12 ms - 20 s (at 99\% confidence level, taking 
into account the number of trials). 

A small H$\alpha$ nebula has been discovered at a position fully consistent
with the coordinates of CXOU J0852 \citep{pellizzoni02}. Such a nebula, if 
physically related to the CCO, could either be a velocity-driven bow-shock
(which would imply that CXOU J0852 is powering a relativistic particle wind), or a 
photo-ionization nebula. Existence of such a diffuse structure was 
confirmed by ESO/VLT observations \citep{mignani07}, which also unveiled
the presence of a faint IR source (Ks$\sim21.4$) close to the position of CXOU J0852. 
However, no firm conclusion about the nature of such source, nor on its
possible association with the CCO, could be drawn. Planned HST observations
in the H$\alpha$ band will shed light on the nature of the diffuse structure.

\subsection{``Candidate'' CCOs}

Few more X-ray sources have been observed inside supernova
remnants, with a phenomenology pointing to a classification
as CCOs. 

Chandra images have unveiled a possible CCO at the center of
the $\sim3000$ yr old SNR G330.2+1.0 \citep{park06}, a member
of the class of non-thermal supernova remnants \citep{torii06},
located at 5-10 kpc, very similar to G347.3-0.5 and Vela Jr.
Such a point source ($\sim600$ counts) shows spectrum and 
luminosity very similar
to other CCOs; a marginal evidence for pulsations at 7.5 s
has also been obtained. 

A possible CCO has been discovered with Chandra
close to the center of the
very young (1000-3000 yr) shell-type 
SNR G15.9+0.2, located at $\sim8.5$ kpc.
The spectrum and luminosity of the point source, 
highly absorbed, together with the lack
of radio or optical counterpart,
seem typical for a CCO, altough a very
small statistics is available ($\sim100$ counts). 

Chandra images unveiled an X-ray source 
close to the center of the 
G349.7+0.2, a $\sim4000$ yr old SNR located at $\sim22$ kpc
 \citep{lazendic05}. The small number of photons ($\sim30$ counts)
hampers any further consideration. However,
if the source
is associated to the SNR, its luminosity would point to 
a CCO interpretation.

RX J0002+6245, an X-ray source located close to the CTA1 
supernova remnant, was proposed by \citet{hailey95} to be an
INS, on the basis of the thermal-like spectrum and possible pulsation
at 242 ms. Faint surrounding diffuse emission  
was proposed to be a previously unknown SNR, associated to the 
INS. XMM-Newton observations do not confirm such a picture and 
clearly show that RX J0002+6245 is a normal F-type star \citep{esposito07}. 

\section{Conclusions}

Sensitive multiwavelength observations point to the picture of CCOs
as an heterogeneous sample of intrinsically different objects.
We are pretty sure that 1E1207 and CXOU J1852 are neutron stars
with a weak magnetic dipole field. On the other side, 1E 1613 is possibly a very
peculiar magnetar, and the central source in Cas A could also be
a magnetar in a long-lasting quiescent phase. 
What could the remaining CCOs be? The birth rate of 
objects like 1E 1613 (be it a braked magnetar, 
or a young binary) is expected to be very low, thus it seems unlikely
to find similar sources hidden (in quiescence?) behind other CCOs.
Most probably, CCOs include both
weakly magnetized INSs and dormant magnetars.
Thus, they  represent the
two wings of the distribution of newborn neutron stars as a function
of their magnetic fields, bracketing the radio pulsars which account for 
the bulk of the population. 
Ironically, our current view
of the CCO phenomenology in several cases prevents us 
from distinguishing 
between
two alternative scenarios requiring totally different 
physical properties. 

The scenario of weakly magnetized INSs 
is based on the link between slow rotation of the 
proto-neutron star, inefficient magnetic field generation and
accretion of fallback material,
which would quench standard ``radio PSR'' emission. 
A sort of unified picture, in which the evolution
of an INS depends on initial spin/magnetic field properties,
driving the star's interaction with fallback material, could be
considered (as suggested a few years ago by \citep{alpar01}).
The biggest problem within 
the weakly magnetized INSs scenario is accounting for
the details of the X-ray emission, explaining
the rich phenomenologies of the 
prototypes 1E 1207 and CXOU J1852 and 
the less spectacular properties of the other candidates. 
Why do we see 
multiple spectral features in 1E 1207 only? How can the pulsed fraction 
in CXOU J1852 be so high? Why the pulsed fraction
of the other sources is so low? 
Where are X-rays ultimately produced?
Are we seeing the neutron star surface? 

A lot of theoretical 
work will be needed. Other interesting issues are the birth rate 
of such weakly magnetized INSs, and their ``fate''. After the
host SNR fades away, such sources,
which were found during observations
devoted to the study of their SNRs, could quickly become much harder
to detect, replenishing
the large expected (but not observed) 
Galactic population of INSs. Or, alternatively, could
they begin at a later stage a radio PSR activity? Sensitive searches
for radio pulsations from 1E 1207 and CXOU J1852 would be very interesting,
especially in view of the possible detection of 1E 1207 as a radio PSR
from Parkes \citep{camilo03}. 

The picture of dormant magnetars is also a viable possibility.
Assessing a magnetar nature for
one or more CCOs would have important consequences on our estimate
of the Galactic population of magnetars (many more could hide in a long-lasting
quiescent state) and of the birth rate of such sources.

Sensitive X-ray (and radio) searches for pulsations and for 
long-term variability,
coupled to deep observations in the infrared (to search for a possible
debris disc - current upper
limits are not constraining) will be crucial to address the nature of CCOs. 
It will be a rewarding investment, since it will
complete our view of the birth properties of neutron stars. This will be a
fundamental piece of information in order to 
derive a coherent, unified scenario for
different species of INSs, elucidating
which differences are related to the objects' nature (birth properties)
and which ones are related to the objects' evolution. 
Indeed, as noted by Woods (these proceedings), the combination of the
estimated birth rates for different INSs species \citep[see, e.g.][for radio
  PSRs, magnetars, RRaTs and XDINSs, respectively]{faucher05,gill07,popov06}
exceeds the overall estimated
Galactic core-collapse supernova rate \citep{diehl06}. Although
such estimates should be taken with caution, this suggests the possibility of 
an evolutionary path linking at least few INSs species. We could also
expect at least few
Galactic SNRs to host an Isolated Black Hole (IBH) and thus
we cannot exclude that some IBH be hidden
among CCOs (as it was considered for the source in Cas A
\citep{pavlov00,chakrabarty01}). Such an hypothesis seems rather
unlikely because there are no IBH emission models able to fit the
observed X-ray emission properties \citep{chakrabarty01}. Furthermore, 
we would be facing some sort of conspiracy, rendering undistinguishable 
the phenomenologies of astrophysical objects as diverse 
as weakly magnetized neutron stars, dormant magnetars and IBHs.


\begin{theacknowledgments}
I warmly thank the organizers of the conference ``40 years of
Pulsars: Millisecond Pulsars, Magnetars and more'' for the invitation. 
I thank P.A.Caraveo and P.Esposito for a critical reading of the manuscript,
and S.Mereghetti, R.P.Mignani, A.Pellizzoni,
A.Tiengo and G.F.Bignami for many useful discussions. 
My research work on the topic of the manuscript is supported by 
the Italian Space Agency (ASI).
\end{theacknowledgments}


\bibliographystyle{aipproc}   




\end{document}